\documentclass[aps,prl,superscriptaddress,twocolumn,nopacs,amsmath,amssymb,letter]{revtex4}

\setcitestyle{super}
\usepackage{color}
\usepackage{graphicx}
\usepackage{dcolumn}
\usepackage{bm}
\usepackage{ulem}

\begin{document}

\title{Spin-valley locking in the normal state of a \\ transition-metal dichalocogenide superconductor}

\author{L.~Bawden}
\affiliation {SUPA, School of Physics and Astronomy, University of
St. Andrews, St. Andrews, Fife KY16 9SS, United Kingdom}

\author{S.~P.~Cooil}
\author{F.~Mazzola}
\affiliation{Department of Physics, Norwegian University of Science and Technology (NTNU), N-7491 Trondheim, Norway}

\author{J.~M.~Riley}
\affiliation {SUPA, School of Physics and Astronomy, University of
St. Andrews, St. Andrews, Fife KY16 9SS, United Kingdom}
\affiliation{Diamond Light Source, Harwell Campus, Didcot, OX11 0DE, United Kingdom}

\author{L.~J.~Collins-McIntyre}
\affiliation {SUPA, School of Physics and Astronomy, University of
St. Andrews, St. Andrews, Fife KY16 9SS, United Kingdom}

\author{V.~Sunko}
\affiliation {SUPA, School of Physics and Astronomy, University of
St. Andrews, St. Andrews, Fife KY16 9SS, United Kingdom}
\affiliation {Max Planck Institute for Chemical Physics of Solids, N{\"o}thnitzer Stra{\ss}e 40, 01217 Dresden, Germany}

\author{K.~Hunvik}
\affiliation{Department of Physics, Norwegian University of Science and Technology (NTNU), N-7491 Trondheim, Norway}

\author{M.~Leandersson}
\author{C.~M.~Polley}
\author{T.~Balasubramanian}
\affiliation{MAX IV Laboratory, Lund University, P. O. Box 118, 221 00 Lund, Sweden}

\author{T.~K.~Kim}
\author{M.~Hoesch}
\affiliation{Diamond Light Source, Harwell Campus, Didcot, OX11 0DE, United Kingdom}

\author{J.~W.~Wells}
\affiliation{Department of Physics, Norwegian University of Science and Technology (NTNU), N-7491 Trondheim, Norway}

\author{G.~Balakrishnan}
\affiliation{Department of Physics, University of Warwick, Coventry CV4 7AL, United Kingdom}

\author{M.~S.~Bahramy}
\affiliation{Quantum-Phase Electronics Center and Department of Applied Physics, The University of Tokyo, Tokyo 113-8656, Japan}
\affiliation{RIKEN center for Emergent Matter Science (CEMS), Wako 351-0198, Japan}

\author{P.~D.~C.~King}
\altaffiliation {To whom correspondence should be addressed: philip.king@st-andrews.ac.uk}
\affiliation {SUPA, School of Physics and Astronomy, University of
St. Andrews, St. Andrews, Fife KY16 9SS, United Kingdom}  

\begin{abstract} The metallic transition-metal dichalcogenides (TMDCs) are benchmark systems for studying and controlling intertwined electronic orders in solids, with superconductivity developing upon cooling from a charge density wave state. The interplay between such phases is thought to play a critical role in the unconventional superconductivity of cuprates, Fe-based, and heavy-fermion systems, yet even for the more moderately-correlated TMDCs, their nature and origins have proved highly controversial. Here, we study a prototypical example, $2H$-NbSe$_2$, by spin- and angle-resolved photoemission and first-principles theory. We find that the normal state, from which its hallmark collective phases emerge, is characterised by quasiparticles whose spin is locked to their valley pseudospin. This results from a combination of strong spin-orbit interactions and local inversion symmetry breaking. Non-negligible interlayer coupling further drives a rich three-dimensional momentum-dependence of the underlying Fermi surface spin texture.  Together, these findings necessitate a fundamental re-investigation of the nature of charge order and superconducting pairing in NbSe$_2$ and related TMDCs.
\end{abstract}

\date{\today}
\maketitle

\newpage

\noindent In combination with broken structural inversion symmetry, spin-orbit coupling (SOC) provides a powerful route to stabilise spin-polarised electronic states without magnetism. This can give rise to electrically-tuneable spin splittings via the Rashba effect, promising new technological developments in spintronics,~\cite{manchon_new_2015} and underpins the formation of spin-helical Dirac cones at the surfaces of topological insulators.~\cite{hasan_topological_2010} It is strongly desired to realise similar effects in systems where more pronounced electronic interactions drive the emergence of collective phases. In non-centrosymmetric superconductors, for example, spin-splitting driven by strong SOC is expected to induce a mixing of spin-triplet and singlet superconducting order parameters,~\cite{gorkov_superconducting_2001} and offers potential for stabilising topological superconductors.~\cite{sato_topological_2009} Yet, identifying suitable candidate materials has proved a major challenge to date. Partly this is driven by a relative dearth of non-centrosymmetric metals, while for non-magnetic systems in which the centre of inversion is maintained, robust spin-degeneracies of their electronic states would typically be expected due to the dual constraints of time-reversal and inversion symmetry.

In contrast, we show here from spin- and angle-resolved photoemission (ARPES) measurements that the normal state of the centrosymmetric TMDC superconductor $2H$-NbSe$_2$ (hereafter denoted NbSe$_2$) hosts a strong layer-resolved and momentum-dependent spin-polarisation of its electronic states at and in the vicinity of the Fermi level. We attribute this as a consequence of a recently-realised form of spin polarisation that can emerge in globally centrosymmetric materials in which constituent structural units nonetheless break inversion symmetry.~\cite{riley_direct_2014,zhang_hidden_2014,sigrist_superconductors_2014} Together with first-principles calculations, we show how this drives a critical and complex interplay of interlayer interactions and spin-orbit coupling in NbSe$_2$. This yields a rich underlying spin-polarised electronic landscape from which charge order and superconductivity emerge upon cooling. 

\begin{figure*}
\begin{center}
\includegraphics[width=\textwidth]{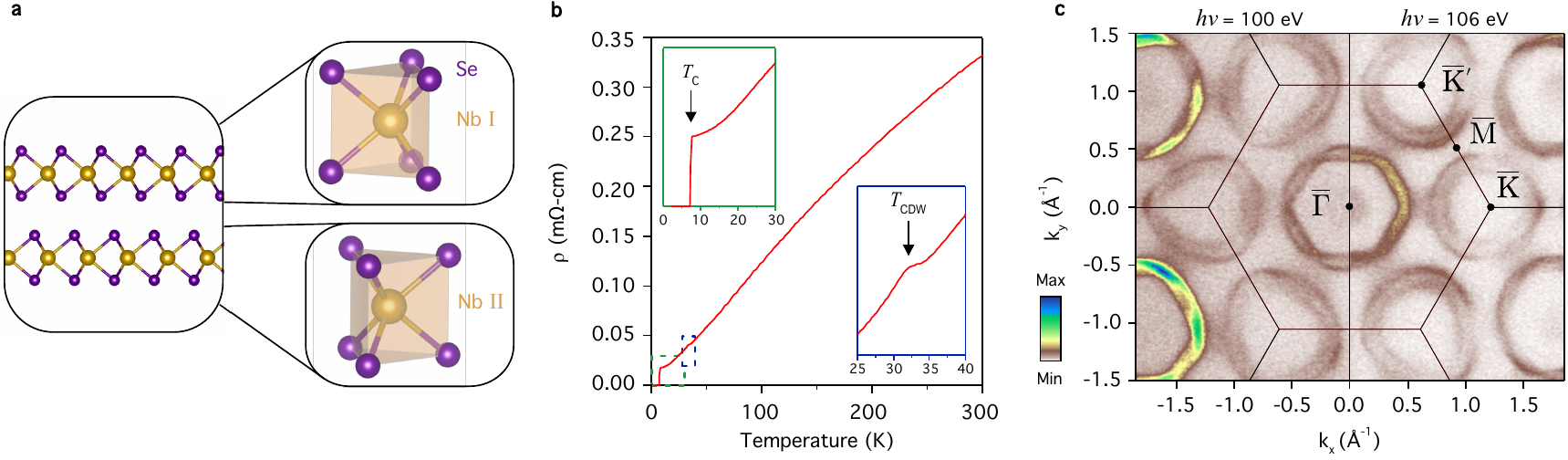}
\caption{ \label{f:overview} {\bf Superconductivity and charge density wave order in $2H$-NbSe$_2$.} (a) Centrosymmetric bulk crystal structure (side view) of $2H$-NbSe$_2$. This is formed by stacking non-centrosymmetric layers of $D_{3h}$ symmetry with 180$^\circ$ relative rotations, restoring the bulk inversion centre. (b) Resistivity measurements show clear signatures of charge-density wave formation at $T_{\mathrm{CDW}}\approx33$~K and superconductivity at $T_{\mathrm{c}}\approx7$~K (magnified in the insets). (c) Normal-state Fermi surface measured by ARPES with $h\nu=100$~eV (left-hand-side) and $h\nu=106$~eV (right-hand-side); $E_F\pm20$~meV. This consists of two Nb-derived barrels centred around the zone-corner $\overline{K}$ ($\overline{K'}$) points, two Nb-derived barrels at the Brillouin zone centre, and an additional central diffuse pocket (most visible at $h\nu=106$~eV) predominantly derived from Se~$p_z$ orbitals.}
\end{center}
\end{figure*}

\

{\noindent \bf Results}

{\noindent \bf Bulk electronic properties of NbSe$_2$.} We consider here exclusively the layered $2H$ polymorph (Fig.~\ref{f:overview}(a)). Each layer forms a graphene-like honeycomb structure with Nb occupying the ``A'' sublattice and two Se atoms situated on the ``B'' sublattice. These lie out of the basel plane, equidistant above and below the transition metal. The unit cell contains two such layers, stacked along the $c$-axis with 180$^\circ$ in-plane rotation. Our resistivity measurements (Fig.~\ref{f:overview}(b)) from single-crystal NbSe$_2$ samples show a metallic temperature-dependence. They additionally exhibit a pronounced hump at a temperature of $T_{\mathrm{CDW}}\approx33$~K indicative of charge-density wave (CDW) formation,\cite{wilson_charge-density_1974} as well as a sharp superconducting transition at $T_{\mathrm{c}}\approx7$~K. The corresponding normal-state Fermi surface is shown in Fig.~\ref{f:overview}(c), as measured by ARPES. There are two barrels centred around each zone-corner $\overline{K}$ ($\overline{K'}$) point which are strongly trigonally-warped. Two further barrels are centred at the $\overline{\Gamma}$ point, the inner of which is hexagonal while the outer exhibits additional warping. From our first-principles density-functional theory calculations, and consistent with previous studies,~\cite{rossnagel_fermi_2001,johannes_fermi-surface_2006,flicker_charge_2015} we assign all four of these Fermi surface sheets as being predominantly derived from Nb $4d$ orbitals. Additional spectral weight at the zone centre is evident in our ARPES measurements for selected photon energies, which we attribute as a fifth, highly three-dimensional, Fermi surface sheet of predominantly Se $p_z$ orbital character. 

\begin{figure*}
\begin{center}
\includegraphics[width=\textwidth]{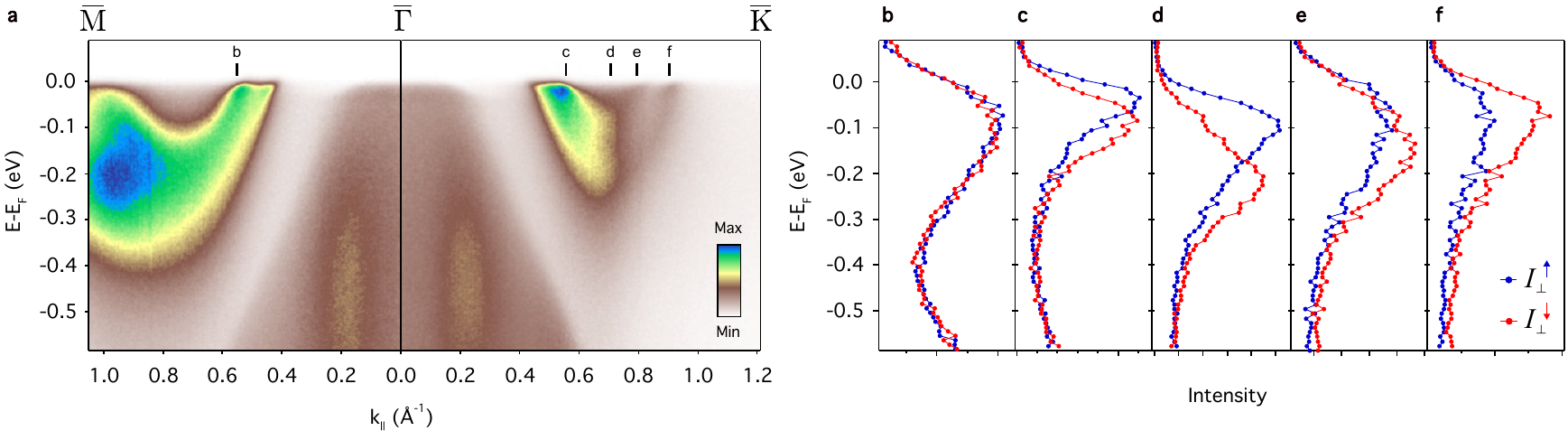}
\caption{ \label{f:GK} {\bf Spin-polarised bulk electronic structure.} (a) Dispersion measured by ARPES ($h\nu=22$~eV) along the $\overline{M}-\overline{\Gamma}-\overline{K}$ direction. (b)-(f) Spin-resolved EDCs at the momenta marked in (a), revealing a strong spin polarisation of these electronic states along $\overline{\Gamma}-\overline{K}$.}
\end{center}
\end{figure*}
\begin{figure}
\begin{center}
\includegraphics[width=\columnwidth]{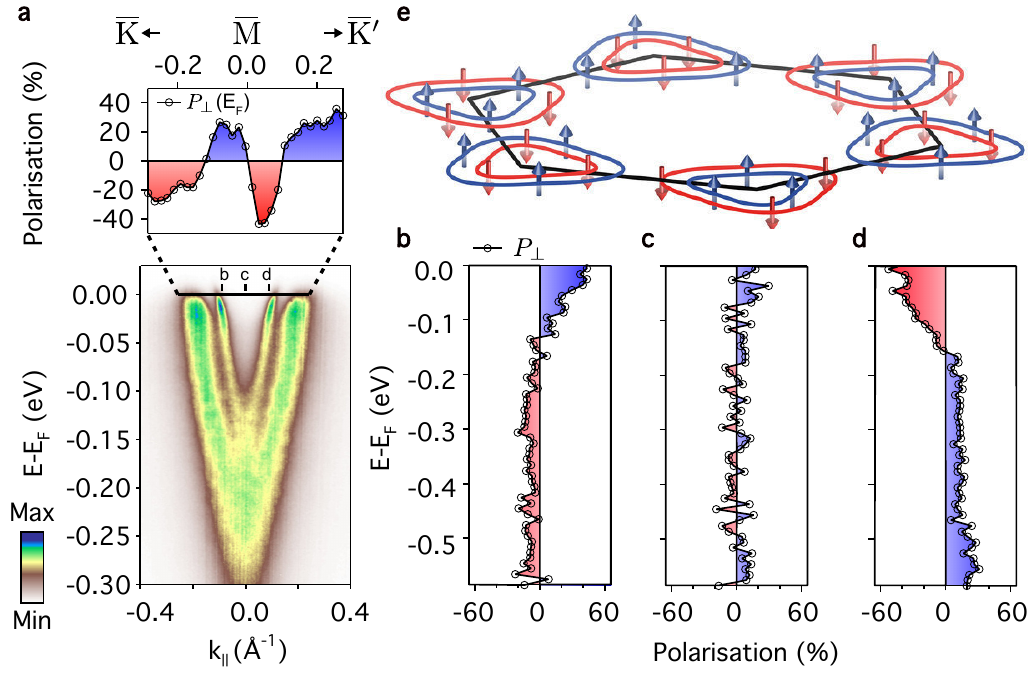}
\caption{ \label{f:KMK} {\bf Spin-valley locked Fermi surfaces.} (a) Dispersion measured ($h\nu=22$~eV) along the $\overline{K}-\overline{M}-\overline{K}'$ direction, together with the corresponding spin-polarisation of an MDC at the Fermi level and (b-d) EDCs at the momenta marked in (a). These reveal how the sign of the spin polarisation for each zone-corner Fermi surface sheet becomes locked to the valley degree of freedom, as shown schematically in (e).}
\end{center}
\end{figure}

This sheet also contributes diffuse filled-in intensity, due to the finite out-of-plane momentum ($k_z$) resolution of ARPES, close to the zone centre in measurements of the electronic dispersions along in-plane high-symmetry directions (Fig.~\ref{f:GK}(a)). The Nb-dominated states, on the other hand, yield clear spectral features close to their Fermi crossings. Kinks in their measured dispersion and a decrease in linewidth near the Fermi level point to relatively-strong electron-phonon coupling in this system.~\cite{Weber_2011,rahn_gaps_2012,flicker_charge_2015} It is these Nb-derived states that are known to host the largest energy gaps at the Fermi level arising from the CDW and superconducting instabilities in this system,~\cite{yokoya_fermi_2001,kiss_charge-order-maximized_2007,borisenko_two_2009,rahn_gaps_2012} although the origins of these have proved highly controversial.~\cite{wilson_charge-density_1974,rice_new_1975,yokoya_fermi_2001,johannes_fermi-surface_2006,kiss_charge-order-maximized_2007,shen_primary_2008,borisenko_two_2009,rossnagel_origin_2011,rahn_gaps_2012,soumyanarayanan_quantum_2013,laverock_$k$-resolved_2013,flicker_charge_2015,xi_strongly_2015} To the best of our knowledge, all prior theoretical treatments assume these orders emerge from an electronic liquid of trivially spin-degenerate character. Indeed, standard expectations of group theory would say this must be the case for the centrosymmetric space group ($P6_3/mmc$) of $2H$-NbSe$_2$.

In contrast, we directly observe pronounced spin polarisations of the underlying electronic states in spin-resolved energy distribution curves (spin-EDCs, see Methods) measured along the $\overline{\Gamma}-\overline{K}$ direction, shown in Fig.~\ref{f:GK}(c-f). This is particularly apparent close to the saddle point of these bands (Fig.~\ref{f:GK}(d)), where two clearly-separated peaks can be observed in spin-EDCs. The measured spin polarisation is almost entirely out-of-plane (Fig.~\ref{f:GK}(d) and Supplemental Fig.~S1), with a sign that reverses between the two bands. We attribute this as arising from local inversion symmetry breaking within the individual layers that make up the bulk crystal structure.~\cite{riley_direct_2014,zhang_hidden_2014,riley_negative_2015} For sufficiently weak interlayer interactions (a point we return to below), pronounced spin-orbit coupling characteristic of the $4d$ transition metal can lift the spin-degeneracy of the states localised within each layer, that thus strongly ``feel'' the local inversion asymmetry. A layer-dependent sign change of this spin polarisation, mediated by the rotated layer stacking of the bulk crystal structure, would act to restore overall spin degeneracy as required by global inversion and time-reversal symmetries. Nonetheless, the electronic states retain strong layer-resolved spin polarisations. Depth-averaging probes would largely be insensitive to these, but photoemission is a surface sensitive technique. Incoherent superposition of photoelectrons emitted from neighbouring layers of the unit cell would lead to some suppression of the measured spin polarisation, while interference effects can further complicate this picture~\cite{riley_direct_2014}. Nonetheless, the extreme surface sensitivity at the photon energies used here, with an inelastic mean free path on the order of the inter-layer separation, renders us predominantly sensitive to the spin texture of the top-most layer of the unit cell.

\

{\noindent \bf Spin-valley locking.} Our measurements show that such non-trivial spin textures in NbSe$_2$ persist up to the Fermi level. This is evident in spin-resolved EDCs along the $\overline{\Gamma}-\overline{K}$ direction (Fig.~\ref{f:GK}(c-f)) as well as the spin polarisation of EDCs and a Fermi-level momentum distribution curve (MDC) along the $\overline{K}-\overline{M}-\overline{K'}$ direction (Fig.~\ref{f:KMK}). The latter clearly reveals how the spin polarisation reverses sign for each pair of Fermi surface sheets centred on neighbouring $\overline{K}$ and $\overline{K'}$ points. This is a natural consequence of time-reversal symmetry. Here, this results in a coupling of the spin to the so-called valley index, the quantum number which distinguishes $\overline{K}$- and $\overline{K'}$-centred Fermi surfaces in NbSe$_2$ (Fig.~\ref{f:KMK}(e)). Such spin-valley coupling has recently been extensively investigated for the band extrema in monolayers of semiconducting TMDCs,~\cite{xiao_coupled_2012,mak_control_2012,zeng_valley_2012} where it has not only been shown to lead to new physics, such as a valley Hall effect,~\cite{mak_valley_2014} but also to offer potential for devices exploiting the valley pseudospin.~\cite{xu_spin_2014,gong_magnetoelectric_2013} Our observations here point to a pronounced role of spin-valley coupling also for the low-energy quasiparticle excitations of the metallic $2H$-structured TMDCs.  

\begin{figure*}
\begin{center}
\includegraphics[width=\textwidth]{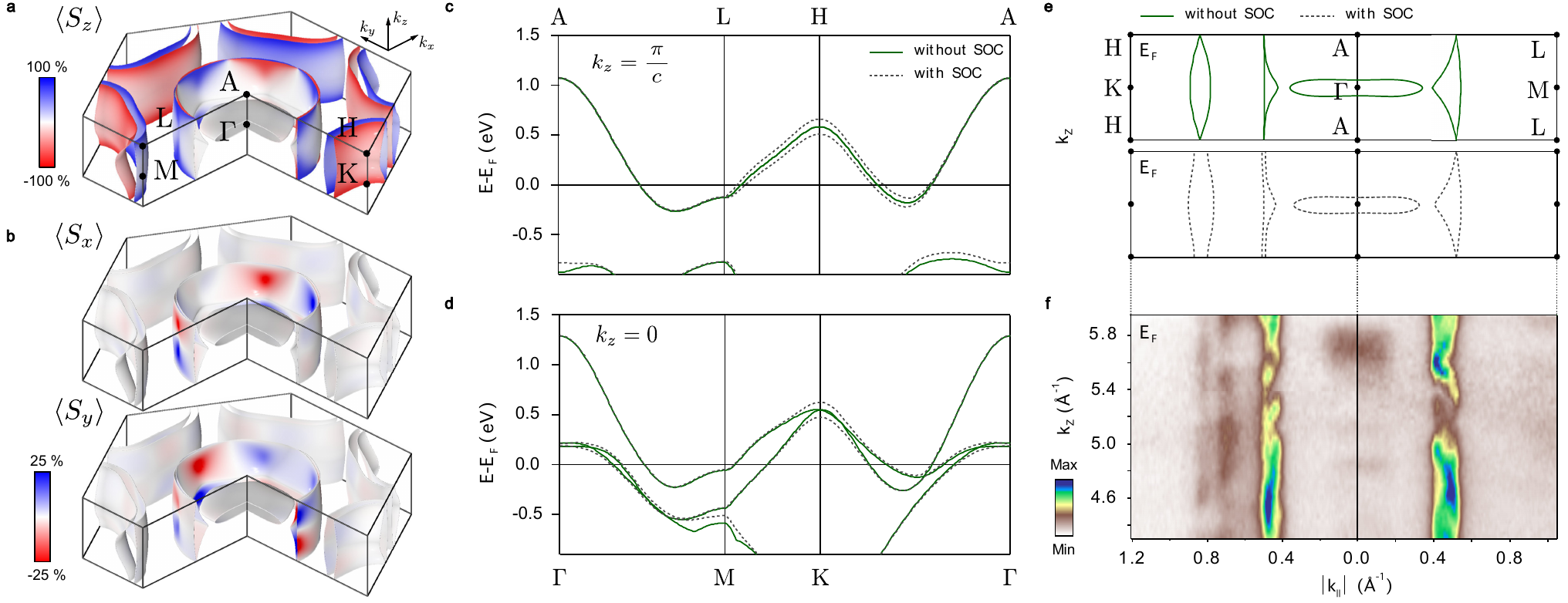}
\caption{ \label{f:DFT} {\bf Interplay of interlayer interactions and intralayer inversion symmetry breaking.} (a,b) Density-functional theory calculations of the (a) out-of-plane and (b) in-plane spin polarisation of the three-dimensional Fermi surface of NbSe$_2$ projected onto the first layer of the unit cell. (c,d) Calculated electronic structure along (c) $A$-$L$-$H$-$A$ ($k_z=\pi/c$) and (d) $\Gamma$-$M$-$K$-$\Gamma$ ($k_z=0$) with and without SOC. (e) Corresponding influence of SOC on the Fermi surface contours in the $\Gamma$-$K$-$H$-$A$ and $\Gamma$-$M$-$L$-$A$ planes (shown throughout the full three-dimensional Brillouin zone in Supplemental Fig.~S2). (f) Our experimental ARPES measurements of such $k_z$-dependent Fermi surfaces ($h\nu=60$ to $130$ eV) are in good general agreement with the theoretical calculations including SOC, supporting a SOC-mediated suppression of interlayer hopping in the $\Gamma$-$K$-$H$-$A$ plane. }
\end{center}
\end{figure*}

Critically, it is from this spin-polarised Fermi sea that electron-hole and electron-electron pairing interactions drive the formation of CDW order and superconductivity. The largest CDW gaps in NbSe$_2$ are located on the zone-corner spin-valley locked Fermi surfaces.~\cite{rahn_gaps_2012} The CDW wave vector is entirely in-plane,~\cite{CDW_neutrons} making such hidden layer-dependent spin polarisations relevant: we thus conclude that these cannot be viewed as purely {\it charge} density wave states with no role of the spin degree of freedom. The underlying spin textures can also be expected to have important implications for superconductivity. Recent measurements on electrically-gated MoS$_2$, which is known to host a strong spin-valley locking in its band structure,~\cite{mak_control_2012,zeng_valley_2012} have found evidence for unconventional so-called Ising superconductivity.~\cite{lu_evidence_2015,saito_superconductivity_2015} In this, the pronounced spin-orbit field that induces the underlying spin texture of the electronic states pins the spin of Cooper pair electrons in the out-of-plane direction. This leads to upper critical fields dramatically exceeding the Pauli paramagnetic limit for a magnetic field applied within the {\textit{ab}}-plane.  Similar phenomenology has recently been reported for mono- and few-layer NbSe$_2$,~\cite{xi_ising_2015} entirely consistent with our direct observation of spin-valley locking in this compound.

In bulk NbSe$_2$, the spin-layer locking, and also multi-band nature as compared to gated MoS$_2$, raises further prospects for stabilising a delicate balance between different pairing states. In the normal state, we find that the $\overline{\Gamma}$-centred Fermi surface barrels, which are known to support a modulated superconducting gap in the bulk,~\cite{rahn_gaps_2012,huang_experimental_2007} are also strongly spin-polarised along the $\overline{\Gamma}-\overline{K}$ direction (Fig.~\ref{f:GK}(c)). Their spin polarisation, however, is completely suppressed along $\overline{\Gamma}-\overline{M}$ (Fig.~\ref{f:GK}(b)). Intriguingly, the largest superconducting gaps of this Fermi surface determined in Ref.~\onlinecite{rahn_gaps_2012} are located at the regions of strongest spin polarisation evident here. Pronounced superconducting gaps are also known to occur on the zone-corner Fermi surfaces,~\cite{rahn_gaps_2012,yokoya_fermi_2001,huang_experimental_2007} which, as demonstrated above, host strong spin-valley locking even in the bulk. Given the significant influence of SOC, a mixing of spin-triplet and spin-singlet order parameters could potentially be expected.~\cite{gorkov_superconducting_2001} Forming even pseudospin-singlet Cooper pairs from the strongly spin-polarised Fermi surfaces could necessitate a phase locking between the order parameter of neighbouring Fermi surface sheets. Moreover, the $c$-axis coherence length in NbSe$_2$ is much greater than the interlayer separation.~\cite{sanchez_specific_1995,nader_critical_2014} This raises the tantalising prospect that the inherent coupling between the spin and layer pseudospins reported here could be tuned to drive an instability to an odd-parity pair density wave state, where the sign of the superconducting gap becomes tied to the layer index.~\cite{sigrist_superconductors_2014,yoshida_pair-density_2012} The proximity of bulk NbSe$_2$ and similar compounds to such phases requires further theoretical exploration, and will depend sensitively on the relative importance of inter- and intra-band as well as interlayer pairing interactions.

\

{\noindent \bf Three-dimensional spin structure.} Our density-functional theory calculations (Fig.~\ref{f:DFT}) already reveal a key role of interlayer interactions in mediating and controlling the underlying spin texture of the normal-state bulk Fermi surface. The calculated spin-polarisation projected onto the first layer of the unit cell is shown throughout the full three-dimensional Brillouin zone in Fig.~\ref{f:DFT}(a,b). The momentum-dependent spin polarisations are determined by the effective spin-orbit field, $B_{so}=\beta(\nabla{V}\times{\mathbf{k})}$, where $\beta$ is a momentum-dependent scaling factor, $\nabla{V}$ is the net electrostatic potential gradient, and $\mathbf{k}$ is the crystal momentum. The horizontal mirror symmetry of each NbSe$_2$ structural unit about the transition-metal plane ($\sigma_h$ of the $D_{3h}$ point group; Fig.~\ref{f:overview}(a)) ensures that $\nabla{V}$ is entirely within the $xy$ plane. Due to the 180$^\circ$ relative rotation of neighbouring NbSe$_2$ monolayers in the bulk crystal structure, $\nabla{V}$, and thus $B_{so}$, has opposite sign for successive layers in the unit cell. This causes the sign of the spin polarisation to reverse at each momentum-space point when projected onto layer 2 vs.\ layer 1 of the unit cell (Supplemental Fig.~S3). This confirms that local inversion asymmetry within each NbSe$_2$ layer drives the formation of the spin-polarised states observed here. 

For electronic states whose wavefunctions are completely delocalised over both layers of the unit cell, the spin-orbit field from each layer cancels. Such states are consequently unpolarised, restoring the conventional expectations for a centrosymmetric space group. This can be observed for the highly three-dimensional ``pancake'' Fermi surface at the Brillouin zone centre here. In contrast, for electronic states predominantly localised within individual layers of the unit cell, strong layer-dependent spin-orbit fields mediate large layer-resolved spin polarisations. This is ideally realised at the three-dimensional Brillouin zone boundary along $k_z$ ($k_z=\pi/c$). In a tight-binding picture, for this $k_z$, interlayer hoppings within a unit cell and between neighbouring unit cells are out of phase with each other and thus cancel.~\cite{johannes_fermi-surface_2006} This can be directly visualised by comparing the dispersion of the electronic states along the $\Gamma$-$M$-$K$-$\Gamma$ and $A$-$L$-$H$-$A$ directions: neglecting SOC, a single four-fold degenerate band crosses the Fermi level in the $k_z=\pi/c$ plane (Fig.~\ref{f:DFT}(c)), whereas this is split into a pair of two-fold degenerate bands by interlayer interactions for $|k_z|<\pi/c$, as evident in Fig.~\ref{f:DFT}(d).

With its forbidden interlayer coupling, the electronic structure for the $k_z=\pi/c$ plane is thus formally equivalent to that of an isolated monolayer. The spin-orbit field is consequently maximised, driving the largest ($>90$\%) layer-dependent spin-polarisations. These are purely out-of-plane (Fig.~\ref{f:DFT}(a,b)), as $(\nabla{V}\times\mathbf{k})$ must lie entirely along $z$ at the Brillouin zone boundary along $k_z$. Along the $A$-$L$ direction, however, a vertical mirror plane in the crystal structure forbids any out-of-plane component of $(\nabla{V}\times\mathbf{k})$. The spin-orbit field must therefore be strictly zero along $A$-$L$. This enforces a touching, and hence spin-degeneracy, of the zone-centre Fermi surface barrels along this direction, which are otherwise strongly spin-polarised throughout the $k_z=\pi/c$ plane.

For other $k_z$ through the Brillouin zone, the spin-orbit field strength can be partially suppressed by finite interlayer coupling. Moreover, the $z$-component of the momentum induces a non-zero component of $(\nabla{V}\times\mathbf{k})$ in the $xy$ plane.  Together, this causes not only the magnitude of the Fermi surface spin polarisation, but also its vectorial spin texture, to develop a strong dependence on both the in- {\it and} out-of-plane momentum (Fig.~\ref{f:DFT}(a,b)). Within the $\Gamma$-$M$-$L$-$A$ plane, the calculated Fermi surface crossings are relatively strongly dispersive in $k_z$ (Fig.~\ref{f:DFT}(e)). This is consistent with our experimental measurements of the $k_z$-dependent Fermi surface (Fig.~\ref{f:DFT}(f)) and points to significant interlayer coupling. Even away from $k_z=\pi/c$, the Fermi surface spin polarisations therefore become strongly suppressed in the entire vicinity of the $\Gamma$-$M$-$L$-$A$ plane. 

For the $\Gamma$-$K$-$H$-$A$ plane, however, we find that the $k_z$ dispersion of the Fermi surface crossings are significantly reduced by the inclusion of spin-orbit coupling in the calculations (Fig.~\ref{f:DFT}(e)), in keeping with our experimental measurements of more two-dimensional Fermi contours for this plane (Fig.~\ref{f:DFT}(f)). This reduction in $k_z$ dispersion is achieved by a lifting, via SOC, of the four-fold degeneracy that would otherwise be present along the $A$-$H$ line, and is accompanied by the emergence of strong spin-polarisation of the Fermi surface crossings for $k_z=\pi/c$. This is in contrast to the $A$-$L$ line, where the energetic degeneracy of the Fermi crossings is protected by their symmetry-enforced spin degeneracy. The reduction of $k_z$-dispersion within the $\Gamma$-$K$-$H$-$A$ plane is equivalent to a spin-orbit-mediated suppression of interlayer hopping here.~\cite{gong_magnetoelectric_2013} This allows relatively strong spin polarisations to be maintained in the vicinity of this plane, with only a moderate suppression of the spin-orbit field strength away from $k_z=\pi/c$. 

For momenta close to the $K$-$H$ line, the in-plane momentum is always much larger than the out-of-plane component. This causes $(\nabla{V}\times{\mathbf{k})}$ to remain predominately aligned along $z$. The spin polarisations of the zone-corner Fermi surfaces are thus largely out-of-plane throughout the full Brillouin zone (Fig.~\ref{f:DFT}(a)). For the Nb-derived zone-centre Fermi surface barrels, however, much stronger (up to $\sim\!25$\%) in-plane components develop (Fig.~\ref{f:DFT}(b)). The in-plane spin texture is largely radial to the Fermi surface (Supplemental Fig.~S4), and switches sign about the $k_z=0$ plane. This indicates a non-zero component of $(\nabla{V}\times\mathbf{k})$ within the plane, with a direction that is tied to the sign of $k_z$. It thus confirms that an in-plane component of the spin texture arises due to finite out-of-plane momentum, where no symmetry constraint exists to cancel the in-plane component of the spin-orbit field for low-symmetry momenta with $0<|k_z|<\pi/c$. The existence of such hidden in-plane spin textures is a unique property of the bulk compound. Indeed, the in-plane spin component goes strictly to zero at the Brillouin zone boundary along $k_z$, enforced by symmetry, and recovering a monolayer-like purely out-of-plane spin texture for $k_z=\pm\pi/c$. The bulk system thus hosts a rich intertwining of spin, orbital, and layer degrees of freedom, mediating a three-dimensional nature of its spin texture that can be expected to further modulate its pairing interactions.~\cite{saito_superconductivity_2015}

\

{\noindent \bf Discussion}

\noindent Taken together, our results show how the combination of interlayer hopping and intra-layer inversion symmetry breaking can lead to particularly rich momentum-dependent spin textures of metallic TMDCs. Suppression of Fermi surface spin polarisation away from the Brillouin zone boundary along $k_z$ will be reduced with decreased interlayer interactions. Indeed, upper critical fields, already known to exceed the Pauli limit for bulk NbSe$_2$,~\cite{foner_upper_1973} are dramatically enhanced in other $2H$-TMDC superconductors as a function of increasing interlayer separation.\cite{klemm_pristine_2015} This points to a susceptibility of the bulk systems to Ising superconductivity similar to that recently observed in isolated monolayers.~\cite{xi_ising_2015} The delicate balance between interlayer hopping and spin-orbit coupling strength in NbSe$_2$ makes this an ideal material for understanding and ultimately controlling the role of layer-dependent spin polarisations on the collective states and phases of transition-metal dichalcogenides.

\

{\small 
\noindent{\bf Methods}\\
\noindent{\bf ARPES:} Spin-resolved ARPES measurements were performed at the I3 beamline of MAX-IV Laboratory, Sweden, and ARPES measurements were performed using the I05 beamline of Diamond Light Source, UK. Measurements were performed at temperatures of $50-80$~K using $p$-polarised synchrotron light. Scienta R4000 hemispherical electron analysers were utilised for all measurements. This was additionally fitted with a mini-Mott detector scheme for the spin-ARPES measurements, configured to simultaneously probe the out-of-plane and in-plane (along the analyser slit direction) component of the photoelectron spin.~\cite{berntsen_spin-_2010} The finite spin-detection efficiency was corrected using a Sherman function of $S=0.17$,~\cite{berntsen_spin-_2010}  and the spin-resolved EDCs determined according to
\[I_i^{\uparrow,\downarrow} = I_i^{tot}(1\pm{P_i})/2\]
with $i=\{\perp,\parallel\}$, $I_i^{tot} = (I_i^+ + I_i^-)$, $I_i^\pm$  the measured intensity on the individual detectors in the Mott scattering chamber, corrected by a relative detector efficiency calibration, and the total spin polarisation,
\[P_i=\frac{(I_i^+-I_i^-)}{S(I_i^++I_i^-)}.\]

\noindent{\bf Calculations:} Density-functional theory (DFT) calculations including spin-orbit coupling were performed using the modified Becke-Johnson exchange potential and Perdew-Burke-Ernzerhof correlation functional implemented in the WIEN2K program.~\cite{wien2k} A 20x20x10 $k$-mesh was employed. The DFT results were downfolded using maximally localised Wannier functions,~\cite{souza,mostofi,kunes} employing Nb $4d$ orbitals and Se $5p$ orbitals as basis functions. The resulting tight-binding Hamiltonian allows a direct extraction of the spin and layer projections of the electronic structure. 

\

\noindent{\bf Acknowledgements}
We thank M.R.~Lees for assistance with the transport measurements and C.~Hooley, P.~Wahl, A.P.~Mackenzie, S.~Lee, and B.~Braunecker for useful discussions. We gratefully acknowledge support from the Engineering and Physical Sciences Research Council, UK (work at St Andrews under Grant Nos.~EP/I031014/1 and EP/M023427/1 and work at Warwick under Grant No.~EP/M028771/1) and the International Max Planck partnership. PDCK acknowledges support from the Royal Society through a University Research Fellowship. MSB was supported by the Grant-in-Aid for Scientific Research (S) (No. 24224009) from the Ministry of Education, Culture, Sports, Science and Technology (MEXT) of Japan. LB, JR, and VS acknowledge studentship funding from EPSRC through grant nos. EP/G03673X/1, EP/L505079/1, and EP/L015110/1, respectively. The experiments at MAX IV Laboratory were made possible through funding from the Swedish Research Council and the Knut and Alice Wallenberg Foundation. We also thank Diamond Light Source for access to beamline I05 (proposal no. SI11383) that contributed to the results presented here.
}



\begin{thebibliography}{45}
\bibitem{manchon_new_2015}
\bibinfo{author} {Manchon, A. Koo, H.~C., Nitta, J., Frolov, S.~M. and Duine, R.~A.},
\newblock \bibinfo{title}{New perspectives for {Rashba} spin-orbit coupling},
\newblock {\it{\bibinfo{journal}{Nature Mater.}}}
 \textbf{\bibinfo{volume}{14}}, \bibinfo{pages}{871-882}
  (\bibinfo{year}{2015}).
  
 \bibitem{hasan_topological_2010}
\bibinfo{author} {Hasan, M.~Z. and Kane, C.~L.},
\newblock \bibinfo{title}{Colloquium: Topological insulators},
\newblock {\it{\bibinfo{journal}{Rev. Mod. Phys.}}}
 \textbf{\bibinfo{volume}{85}}, \bibinfo{pages}{3045}
  (\bibinfo{year}{2010}).
  
 \bibitem{gorkov_superconducting_2001}
\bibinfo{author} {Gor'kov, L.~P. and Rashba, E.~I.},
\newblock \bibinfo{title}{Superconducting 2D {system} with {lifted} {spin} {degeneracy}: {Mixed} {singlet}-{triplet} {state}},
\newblock {\it{\bibinfo{journal}{Phys. Rev. Lett.}}}
 \textbf{\bibinfo{volume}{87}}, \bibinfo{pages}{037004}
  (\bibinfo{year}{2001}).
  
 \bibitem{sato_topological_2009}
\bibinfo{author}{Sato, M. and Fujimoto, S.},
\newblock \bibinfo{title}{Topological phases of noncentrosymmetric superconductors: {Edge} states, {Majorana} fermions, and non-{Abelian} statistics},
\newblock {\it{\bibinfo{journal}{Phys. Rev. B}}}
 \textbf{\bibinfo{volume}{79}}, \bibinfo{pages}{094504}
  (\bibinfo{year}{2009}).

\bibitem{riley_direct_2014}
\bibinfo{author} {Riley, J.~M.} {\it{et~al.}},
\newblock \bibinfo{title}{Direct observation of spin-polarized bulk bands in an inversion-symmetric semiconductor},
\newblock {\it{\bibinfo{journal}{Nature Phys.}}}
 \textbf{\bibinfo{volume}{10}}, \bibinfo{pages}{835-839}
  (\bibinfo{year}{2014}).

  \bibitem{zhang_hidden_2014}
\bibinfo{author}{Zhang, X., Liu, Q., Luo, J.-W., Freeman, A.~J., and Zunger, A.},
\newblock \bibinfo{title}{Hidden spin polarization in inversion-symmetric bulk crystals},
\newblock {\it{\bibinfo{journal}{Nature Phys.}}}
 \textbf{\bibinfo{volume}{10}}, \bibinfo{pages}{387-393}
  (\bibinfo{year}{2014}).
  
 \bibitem{sigrist_superconductors_2014}
\bibinfo{author}{Sigrist, M.} {\it{et~al.}},
\newblock \bibinfo{title}{Superconductors with {staggered} {non}-centrosymmetricity},
\newblock {\it{\bibinfo{journal}{J. Phys. Soc. Jpn.}}}
 \textbf{\bibinfo{volume}{83}}, \bibinfo{pages}{061014}
  (\bibinfo{year}{2014}).
  
  \bibitem{wilson_charge-density_1974}
\bibinfo{author}{Wilson, J.~A., Di Salvo, F.~J., and Mahajan, S.},
\newblock \bibinfo{title}{Charge-density waves in metallic, layered, transition-metal dichalcogenides},
\newblock {\it{\bibinfo{journal}{Phys. Rev. Lett.}}}
 \textbf{\bibinfo{volume}{32}}, \bibinfo{pages}{882}
  (\bibinfo{year}{1974}).
  
   \bibitem{rossnagel_fermi_2001}
\bibinfo{author}{Rossnagel, K.} {\it{et~al.}},
\newblock \bibinfo{title}{Fermi surface of 2$H$-{NbSe}${_2}$ and its implications on the charge-density-wave mechanism},
\newblock {\it{\bibinfo{journal}{Phys. Rev. B}}}
 \textbf{\bibinfo{volume}{64}}, \bibinfo{pages}{235119}
  (\bibinfo{year}{2001}).
 
\bibitem{johannes_fermi-surface_2006}
\bibinfo{author}{Johannes, M.~D, Mazin, I.~I, and Howells, C.~A.},
\newblock \bibinfo{title}{Fermi-surface nesting and the origin of the charge-density wave in 2$H$-NbSe${_2}$},
\newblock {\it{\bibinfo{journal}{Phys. Rev. B}}}
 \textbf{\bibinfo{volume}{73}}, \bibinfo{pages}{205102}
  (\bibinfo{year}{2006}).
  
\bibitem{flicker_charge_2015}
\bibinfo{author}{Flicker, F. and van Wezel, J.},
\newblock \bibinfo{title}{Charge order from orbital-dependent coupling evidenced by NbSe${_2}$},
\newblock {\it{\bibinfo{journal}{Nature Commun.}}}
 \textbf{\bibinfo{volume}{6}}, \bibinfo{pages}{7034}
  (\bibinfo{year}{2015}).

  \bibitem{Weber_2011}
\bibinfo{author}{Weber, F.} {\it{et~al.}},
\newblock \bibinfo{title}{Electron-phonon coupling and the soft phonon mode in TiSe$_2$},
\newblock {\it{\bibinfo{journal}{Phys. Rev. Lett}}}
 \textbf{\bibinfo{volume}{107}}, \bibinfo{pages}{107403}
  (\bibinfo{year}{2011}).
  
\bibitem{rahn_gaps_2012}
\bibinfo{author}{Rahn, D.~J.} {\it{et~al.}},
\newblock \bibinfo{title}{Gaps and kinks in the electronic structure of the superconductor 2$H$-NbSe${_2}$ from angle-resolved photoemission at $1$~{K}}
\newblock {\it{\bibinfo{journal}{Phys. Rev. B}}}
 \textbf{\bibinfo{volume}{85}}, \bibinfo{pages}{224532}
  (\bibinfo{year}{2012}).

\bibitem{kiss_charge-order-maximized_2007}
\bibinfo{author}{Kiss, T.} {\it{et~al.}},
\newblock \bibinfo{title}{Charge-order-maximized momentum-dependent superconductivity},
\newblock {\it{\bibinfo{journal}{Nature Phys.}}}
 \textbf{\bibinfo{volume}{3}}, \bibinfo{pages}{720-725}
  (\bibinfo{year}{2007}).
  
   \bibitem{yokoya_fermi_2001}
\bibinfo{author}{Yokoya, T., Kiss, T., Chainani, A., Shin, S., Nohara, M. and Takagi, H.},
\newblock \bibinfo{title}{Fermi {surface} {sheet}-{dependent} {superconductivity} in 2$H$-NbSe${_2}$},
\newblock {\it{\bibinfo{journal}{Science}}}
 \textbf{\bibinfo{volume}{294}}, \bibinfo{pages}{2518-2520}
  (\bibinfo{year}{2001}).
  
  \bibitem{borisenko_two_2009}
\bibinfo{author}{Borisenko, S.~V.} {\it{et~al.}},
\newblock \bibinfo{title}{Two {energy} {gaps} and {Fermi}-{surface} ``{arcs}'' in {NbSe}${_2}$},
\newblock {\it{\bibinfo{journal}{Phys. Rev. Lett.}}}
 \textbf{\bibinfo{volume}{102}}, \bibinfo{pages}{166402}
  (\bibinfo{year}{2009}).


  \bibitem{xi_strongly_2015}
\bibinfo{author} {Xi, X.} {\it{et~al.}},
\newblock \bibinfo{title}{Strongly enhanced charge-density-wave order in monolayer {NbSe}${_2}$},
\newblock {\it{\bibinfo{journal}{Nature Nano.}}}
 \textbf{\bibinfo{volume}{10}}, \bibinfo{pages}{765-769}
  (\bibinfo{year}{2015}).

  \bibitem{rice_new_1975}
\bibinfo{author}{Rice, T. M. and Scott, G. K.},
\newblock \bibinfo{title}{New {mechanism} for a {charge}-{density}-{wave} {instability}},
\newblock {\it{\bibinfo{journal}{Phys. Rev. Lett.}}}
 \textbf{\bibinfo{volume}{35}}, \bibinfo{pages}{120}
  (\bibinfo{year}{1975}).
  
\bibitem{shen_primary_2008}
\bibinfo{author}{Shen, D.~W.} {\it{et~al.}},
\newblock \bibinfo{title}{Primary {role} of the {barely} {occupied} {states} in the {charge} {density} {wave} {formation} of {NbSe}${_2}$},
\newblock {\it{\bibinfo{journal}{Phys. Rev. Lett.}}}
 \textbf{\bibinfo{volume}{101}}, \bibinfo{pages}{226406}
  (\bibinfo{year}{2008}).

\bibitem{rossnagel_origin_2011}
\bibinfo{author}{Rossnagel, K.},
\newblock \bibinfo{title}{On the origin of charge-density waves in select layered transition-metal dichalcogenides},
\newblock {\it{\bibinfo{journal}{J. Phys. Cond. Mat.}}}
 \textbf{\bibinfo{volume}{23}}, \bibinfo{pages}{21}
  (\bibinfo{year}{2011}).
  
 \bibitem{soumyanarayanan_quantum_2013}
\bibinfo{author}{Soumyanarayanan, A.} {\it{et~al.}},
\newblock \bibinfo{title}{Quantum phase transition from triangular to stripe charge order in {NbSe}$_2$}
\newblock {\it{\bibinfo{journal}{P. Nat. Acad. Sci. USA}}}
 \textbf{\bibinfo{volume}{110}}, \bibinfo{pages}{1623-1627}
  (\bibinfo{year}{2013}).
  
  \bibitem{laverock_$k$-resolved_2013}
\bibinfo{author}{Laverock, J.} {\it{et~al.}},
\newblock \bibinfo{title}{$k$-resolved susceptibility function of 2$H$-TaSe$_2$ from angle-resolved photoemission},
\newblock {\it{\bibinfo{journal}{Phys. Rev. B}}}
 \textbf{\bibinfo{volume}{88}}, \bibinfo{pages}{035108}
  (\bibinfo{year}{2013}).

   \bibitem{riley_negative_2015}
\bibinfo{author}{Riley, J.~M.} {\it{et~al.}},
\newblock \bibinfo{title}{Negative electronic compressibility and tunable spin splitting in {WSe}${_2}$},
\newblock {\it{\bibinfo{journal}{Nature Nano.}}}
 \textbf{\bibinfo{volume}{10}}, \bibinfo{pages}{1043-1047}
  (\bibinfo{year}{2015}).

  \bibitem{xiao_coupled_2012}
\bibinfo{author}{Xiao, D., Liu, G.-B., Feng, W., Xu, X. and Yao, W.},
\newblock \bibinfo{title}{Coupled {spin} and {valley} {physics} in {monolayers} of {MoS}${_2}$ and {other} {group}-{VI} {dichalcogenides}},
\newblock {\it{\bibinfo{journal}{Phys. Rev. Lett.}}}
 \textbf{\bibinfo{volume}{108}}, \bibinfo{pages}{196802}
  (\bibinfo{year}{2012}).
  
 \bibitem{mak_control_2012}
\bibinfo{author}{Mak, K.~F., He, K., Shan, J. and Heinz, T.~F.},
\newblock \bibinfo{title}{Control of valley polarization in monolayer {MoS}${_2}$ by optical helicity}
\newblock {\it{\bibinfo{journal}{Nature Nano.}}}
 \textbf{\bibinfo{volume}{7}}, \bibinfo{number}{494-498}
  (\bibinfo{year}{2012}).
  
\bibitem{zeng_valley_2012}
\bibinfo{author}{Zeng, H., Dai, J., Yao, W., Xiao, D. and Cui, X.},
\newblock \bibinfo{title}{Valley polarization in {MoS}${_2}$ monolayers by optical pumping},
\newblock {\it{\bibinfo{journal}{Nature Nano.}}}
\textbf{\bibinfo{volume}{7}}, \bibinfo{number}{490-493}
  (\bibinfo{year}{2012}).

 \bibitem{mak_valley_2014}
\bibinfo{author} {Mak, K.~F., McGill, K.~L., Park, J., and McEuen, P.~L.},
\newblock \bibinfo{title}{The valley {Hall} effect in {MoS}${_2}$ transistors},
\newblock {\it{\bibinfo{journal}{Science}}}
 \textbf{\bibinfo{volume}{344}}, \bibinfo{pages}{1489-1492}
  (\bibinfo{year}{2014}).
  
  \bibitem{gong_magnetoelectric_2013}
\bibinfo{author}{Gong, Z.} {\it{et~al.}},
\newblock \bibinfo{title}{Magnetoelectric effects and valley-controlled spin quantum gates in transition metal dichalcogenide bilayers},
\newblock {\it{\bibinfo{journal}{Nature Comm.}}}
 \textbf{\bibinfo{volume}{4}}, \bibinfo{pages}{2053}
  (\bibinfo{year}{2013}).

  \bibitem{xu_spin_2014}
\bibinfo{author} {Xu, X, Yao, W, Xiao, D. and Heinz, T.~F.},
\newblock \bibinfo{title}{Spin and pseudospins in layered transition metal dichalcogenides},
\newblock {\it{\bibinfo{journal}{Nature Phys.}}}
 \textbf{\bibinfo{volume}{10}}, \bibinfo{pages}{343-350}
  (\bibinfo{year}{2014}).
  
\bibitem{CDW_neutrons}
\bibinfo{author}{Moncton, D.~E.} {\it{et~al.}},
\newblock \bibinfo{title}{Neutron scattering study of the charge-density wave transitions in $2H$-TaSe$_2$ and $2H$-NbSe$_2$},
\newblock {\it{\bibinfo{journal}{Phys. Rev. B}}}
 \textbf{\bibinfo{volume}{16}}, \bibinfo{pages}{801}
  (\bibinfo{year}{1977}).

\bibitem{saito_superconductivity_2015}
\bibinfo{author}{Saito, Y.} {\it{et~al.}},
\newblock \bibinfo{title}{Superconductivity protected by spin-valley locking in gate-tuned {MoS}${_2}$},
\newblock {\it{\bibinfo{journal}{Nature Phys.}}}
\textit{\bibinfo{volume}{12}}, \bibinfo{pages}{144-149}
  (\bibinfo{year}{2016}).

\bibitem{lu_evidence_2015}
\bibinfo{author}{Lu, J.~M.} {\it{et~al.}},
\newblock \bibinfo{title}{Evidence for two-dimensional {Ising} superconductivity in gated {MoS}${_2}$},
\newblock {\it{\bibinfo{journal}{Science}}}
 \textbf{\bibinfo{volume}{350}}, \bibinfo{pages}{1353-1357}
  (\bibinfo{year}{2015}).
  
  \bibitem{xi_ising_2015}
\bibinfo{author} {Xi, X.} {\it{et~al.}},
\newblock \bibinfo{title}{Ising pairing in superconducting {NbSe}$_2$ atomic layers},
\newblock {\it{\bibinfo{journal}{Nature Phys.}}}
 \textit{\bibinfo{volume}{12}}, \bibinfo{pages}{139-143}
  (\bibinfo{year}{2016}).

\bibitem{huang_experimental_2007}
\bibinfo{author}{Huang, C.~L.} {\it{et~al.}},
\newblock \bibinfo{title}{Experimental evidence for a two-gap structure of superconducting NbSe$_2$: A specific-heat study in external magnetic fields},
\newblock {\it{\bibinfo{journal}{Phys. Rev. B}}}
 \textbf{\bibinfo{volume}{76}}, \bibinfo{pages}{212504}
  (\bibinfo{year}{2007}).
  
 \bibitem{sanchez_specific_1995}
\bibinfo{author}{Sanchez, D., Junod, A., Muller, J., Berger, H. and L\'{e}vy, F.},
\newblock \bibinfo{title}{Specific heat of 2$H$-NbSe$_2$ in high magnetic fields},
\newblock {\it{\bibinfo{journal}{Physica B}}}
 \textbf{\bibinfo{volume}{204}}, \bibinfo{pages}{167-175}
  (\bibinfo{year}{1995}).
  
  \bibitem{nader_critical_2014}
\bibinfo{author}{Nader, A. and Monceau, P.},
\newblock \bibinfo{title}{Critical field of 2$H$-NbSe$_2$ down to $50$~mK},
\newblock {\it{\bibinfo{journal}{SpringerPlus}}}
 \textbf{\bibinfo{volume}{3}}, \bibinfo{pages}{16}
  (\bibinfo{year}{2014}).
  
  \bibitem{yoshida_pair-density_2012}
\bibinfo{author}{Yoshida, T., Sigrist, M. and Yanase, Y.},
\newblock \bibinfo{title}{Pair-density wave states through spin-orbit coupling in multilayer superconductors},
\newblock {\it{\bibinfo{journal}{Phys. Rev. B}}}
 \textbf{\bibinfo{volume}{86}}, \bibinfo{pages}{134514}
  (\bibinfo{year}{2012}).  
 
    \bibitem{foner_upper_1973}
\bibinfo{author}{Foner, S. and McNiff Jr, E.~J.},
\newblock \bibinfo{title}{Upper critical fields of layered superconducting NbSe${_2}$ at low temperature},
\newblock {\it{\bibinfo{journal}{Phys. Lett. A}}}
 \textbf{\bibinfo{volume}{45}}, \bibinfo{pages}{429-430}
  (\bibinfo{year}{1973}).
  
 \bibitem{klemm_pristine_2015}
\bibinfo{author}{Klemm, R.~A.},
\newblock \bibinfo{title}{Pristine and intercalated transition metal dichalcogenide superconductors},
\newblock {\it{\bibinfo{journal}{Physica C}}}
\textbf{\bibinfo{volume}{514}}, \bibinfo{pages}{86-94}
  (\bibinfo{year}{2015}). 
  
\bibitem{berntsen_spin-_2010}
\bibinfo{author}{Berntsen, M.~H.} {\it et al.},
\newblock \bibinfo{title}{A spin- and angle-resolving photoelectron spectrometer},
\newblock {\it{\bibinfo{journal}{Rev. Sci. Instrum.}}}
 \textbf{\bibinfo{volume}{81}}, \bibinfo{pages}{035104}
  (\bibinfo{year}{2010}).

\bibitem{wien2k}
Blaha, P. {\it et al.}, WIEN2K package, Version 10.1 (2010); available at http://www.wien2k.at.

\bibitem{souza}
\bibinfo{author}{Souza, I.} {\it et al.},
\newblock \bibinfo{title}{Maximally localized Wannier functions for entangled energy bands},
\newblock {\it{\bibinfo{journal}{Phys. Rev. B}}} \textbf{\bibinfo{volume}{65}},
  \bibinfo{pages}{035109} (\bibinfo{year}{2001}).  

\bibitem{mostofi} 
\bibinfo{author}{Mostofi, A. A.} {\it et al.},
\newblock \bibinfo{title}{Wannier90: A Tool for obtaining maximally localised Wannier functions},
\newblock {\it{\bibinfo{journal}{Comp. Phys. Commun.}}}
\textbf{\bibinfo{volume}{178}}, 
\bibinfo{pages}{685Ð699} (\bibinfo{year}{2008}).
 
   \bibitem{kunes}
\bibinfo{author}{Kune\v s, J.} {\it et al.},
\newblock \bibinfo{title}{WIEN2WANNIER: From linearized augmented plane waves to maximally localized Wannier functions},
\newblock {\it{\bibinfo{journal}{Comp. Phys. Commun. }}} \textbf{\bibinfo{volume}{181}},
  \bibinfo{pages}{1888-1895} (\bibinfo{year}{2010}). 


\end{thebibliography}

\end{document}